# Gate-tunable spin waves in antiferromagnetic atomic bilayers

Xiao-Xiao Zhang[1,2], Lizhong Li[3], Daniel Weber[4], Joshua Goldberger[4], Kin Fai Mak[1,3,5*], Jie Shan[1,3,5*]

[1]Kavli Institute at Cornell for Nanoscale Science, Ithaca, NY 14853, USA
[2]Department of Physics, University of Florida
[3]School of Applied and Engineering Physics, Cornell University, Ithaca, NY 14853, USA
[4]Department of Chemistry and Biochemistry, Ohio State University
[5]Laboratory of Atomic and Solid State Physics, Cornell University, Ithaca, NY 14853, USA.

**The emergence of two-dimensional (2D) layered magnetic materials has opened an exciting playground for both fundamental studies of magnetism in 2D and explorations of spin-based applications [1-4]. Remarkable properties, including spin filtering in magnetic tunnel junctions and gate control of magnetic states, have recently been demonstrated in 2D magnetic materials [5-12]. While these studies focus on the static properties, dynamic magnetic properties such as excitation and control of spin waves have remained elusive. Here we excite spin waves and probe their dynamics in antiferromagnetic $CrI_3$ bilayers by employing an ultrafast optical pump/magneto-optical Kerr probe technique. We identify sub-terahertz magnetic resonances under an in-plane magnetic field, from which we determine the anisotropy and interlayer exchange fields and the spin damping rates. We further show tuning of antiferromagnetic resonances by tens of gigahertz through electrostatic gating. Our results shed light on magnetic excitations and spin dynamics in 2D magnetic materials, and demonstrate their unique potential for applications in ultrafast data storage and processing.**

Spin waves, first predicted by F. Bloch in 1929, are propagating disturbances in magnetic ordering in a magnetic material [13]. The quanta of spin waves are called magnons. The rich spin-wave phenomena in magnetic materials have attracted fundamental interest and impacted on technology of telecommunication systems, radars, and potentially also low-power information transmission and processing due to their decoupling from charge current [14,15]. The main magnetic materials of interest have so far been ferromagnets (FM). The operation speed of FM-based devices is typically in the GHz range, which is limited by the ferromagnetic resonance (zero-momentum resonance) frequency. One of the major attractions of antiferromagnets (AFM), a class of much more common magnetic materials, is the prospect of high-speed operation. The antiferromagnetic resonances are in the frequency range of as high as THz due to the spin-sublattice exchange [16]. The AFMs, however, are difficult to access due to the absence of macroscopic magnetization.

The recent discovery of two-dimensional (2D) layered magnetic materials [17-19], particularly A-type AFMs such as bilayer $CrI_3$ that are made of two antiferromagnetically coupled ferromagnetic monolayers [17], presents new opportunities to unlock the properties of AFMs. With fully uncompensated ferromagnetic surfaces, the magnetic state can be easily accessed and controlled [20]. The van der Waals nature allows their convenient integration into heterostructures with high-quality interfaces [21]. And the atomic thickness allows the application of strong electric field and large electrostatic doping to control the properties of 2D magnetic materials. Although rapid progress has been made in both fundamental understanding and potential applications [1-12,22], the spin dynamics, including basic properties such as magnetic resonances and damping, have remained unexplored in these materials. A major technical challenge arises from the small amount of spins present in atomically thin samples of typical lateral dimensions of a few microns. This makes studies with conventional probes, such as neutron scattering and microwave absorption [23,24], extremely difficult or impossible. Microwave absorption measurements are further hindered by the high antiferromagnetic resonance frequencies.

Here we investigate spin-wave excitations in bilayer $CrI_3$ using the time-resolved magneto-optical Kerr effect (MOKE). The sample consists of a heterostructure of bilayer $CrI_3$ and monolayer $WSe_2$, which is encapsulated in two hexagonal boron nitride (hBN) thin layers for protection of air-sensitive $CrI_3$ (Fig. 1a). While monolayer $CrI_3$ is a ferromagnetic semiconductor with out-of-plane anisotropy below the Curie temperature of about 40 K, bilayer $CrI_3$ is an AFM with spins in the two ferromagnetic monolayers anti-aligned below the Néel temperature of about 45 K [17]. Monolayer $WSe_2$ is a direct gap non-magnetic semiconductor with strong spin-orbit interaction [25]. It is believed to have a type-II band alignment with $CrI_3$ [26] (Fig. 1b). The introduction of $WSe_2$ significantly enhances optical absorption of the pump and hot carrier injection into $CrI_3$ for magnetic excitations. As will be discussed below, $WSe_2$ also breaks the layer symmetry in bilayer $CrI_3$ to enable the detection of different oscillation modes of spin waves in the polar MOKE geometry. Figure 1c is the magnetization of bilayer $CrI_3$ as a function of out-of-plane magnetic field at 4 K probed by magnetic circular dichroism (MCD) at 1.8 eV. The antiferromagnetic behavior is fully consistent with the reported results [17]. The small nonzero magnetization near zero field is a manifestation of the broken layer symmetry. The sharp turn-on of the magnetization around 0.75 T corresponds to a spin-flip transition, which provides a measure of the interlayer exchange field $H_E$.

A pulsed laser (200-fs pulse duration) was employed for the time-resolved measurements. The heterostructure was excited by a light pulse centered near the $WSe_2$ fundamental exciton resonance energy (1.73 eV), and the change in $CrI_3$ magnetization is probed by a time-synchronized pulse at a lower energy (1.54 eV). Both the pump and probe were linearly polarized and at normal incidence. The polarization rotation of the reflected probe beam locked to the modulation frequency of the pump was detected. In this configuration the MOKE signal is sensitive only to the out-of-plane magnetization. An in-plane magnetic field $H_\parallel$ was applied, which causes the magnetization of both the top and bottom layers to cant (Fig. 1d). The field required to rotate the ordered moments into the in-plane direction, which is referred to as the saturation field $H_S$, has been reported to be near 3.8 T for bilayer $CrI_3$ at 2 K [8]. Unless otherwise specified, all measurements were performed at 1.7 K. (See Methods for details on the sample fabrication and the time-resolved MOKE setup.)

Figure 2a displays the time evolution of the pump-induced change in the MOKE signal of bilayer $CrI_3$ under $H_\parallel$ ranging from 0 – 6 T. For all fields, the MOKE signal shows a sudden change at time zero, followed by a decay on the scale of 10's – 100's ps. This reflects the incoherent demagnetization process, in which the magnetic order is disturbed instantaneously by the pump pulse and is slowly reestablished. Oscillations in the MOKE signal that are also instantaneous with the optical excitation become clearly observable with increasing magnetic field. The amplitude, frequency and damping of these oscillations evolve systematically with $H_∥$.

Figure 2b is the fast Fourier transform (FFT) of the oscillatory part of the time traces in Fig. 2a. Two examples are shown in Fig. 3a and 3b for $H_∥$ at 1.5 and 3.75 T, respectively. The exponential decay of the incoherent demagnetization dynamics has been subtracted from the MOKE signal before performing FFT. At low magnetic field, a resonance around 70 GHz is observed. As $H_∥$ increases, it splits into two resonances with one that redshifts significantly and the other that exhibits minimal shifts in frequency until 3.3 T. Above this field, both resonances blueshift with increasing $H_∥$. While the low-energy mode quickly becomes too weak to be observed, the amplitude of the high-energy mode does not depend strongly on field.

We performed a careful analysis of the MOKE dynamics directly in the time domain, fittting the oscillations with two damped harmonic waves (red lines, Fig. 3a, b). The extracted resonance frequencies, damping rates and amplitudes as a function of $H_∥$ are summarized in Fig. 3d, 3e and Supplementary Fig. S7, respectively. We first focus on the resonance frequencies. The field dependence of the resonance frequencies shows two distinct regimes. Below about 3.3 T, the two nearly degenerate modes (at small fields) both soften with increasing field, one slightly and the other nearly to zero frequency. Above 3.3 T, both modes show a linear increase in frequency with a slope equal to the electron gyromagnetic ratio $γ/2π ≈ 28$ GHz/T. The latter is characteristic of a ferromagnetic resonance under high fields.

The observed magnetic-field dispersion of the resonances is indicative of their magnon origin with 3.3 T corresponding to the saturation field $H_S$ in bilayer $CrI_3$. The two modes are the spin precession eigenmodes of the coupled top and bottom layer magnetizations under an in-plane field (Fig. 3c). Above the saturation field, the spins are aligned along the applied field and the spin waves become ferromagnetic-like. This interpretation is further supported by the temperature dependence of the resonances (Supplementary Fig. S3-5). Clear mode softening is observed with increasing temperature and the resonance feature disappears near the Néel temperature of bilayer $CrI_3$. The microscopic mechanism for the observed ultrafast excitation of spin waves in bilayer $CrI_3$ is not fully understood. A plausible process is the exciton generation in $WSe_2$ by the optical pump, followed by ultrafast exciton dissociation and charge transfer at the $CrI_3$-$WSe_2$ interface [26], and an impulsive perturbation to the magnetic interactions [27,28] in $CrI_3$ by the hot carriers. Details on the supporting experiments of this mechanism are provided in Methods.

We model the field dependent spin dynamics using the coupled Landau-Lifshitz-Gilbert (LLG) equations, which describe precession of antiferromagnetically coupled top and bottom layer magnetizations under an in-plane field $H_∥$ [29] (Details are provided in Methods). The effective magnetic field responsible for spin precession in each layer includes contributions from the applied field $H_∥$, intralayer anisotropy field $H_A$, and the interlayer exchange field $H_E$. In the simple case of negligible damping and symmetric top and bottom layers, the frequency of the precession eigenmodes are found as $ω_T = γ\left[H_A(2H_E + H_A) + \frac{2H_E - H_A}{2H_E + H_A} H_∥^2\right]^{\frac{1}{2}}$, $ω_L = γ\left[H_A(2H_E + H_A) - \frac{H_A}{2H_E + H_A} H_∥^2\right]^{\frac{1}{2}}$ (before saturation); and $ω_T = γ\sqrt{H_∥(H_∥ - H_A)}$, $ω_L =$ $\gamma\sqrt{(H_\parallel - 2H_E)(H_\parallel - 2H_E - H_A)}$ (after saturation). As shown schematically in Fig. 3c, the low-energy mode corresponds to the longitudinal (with respect to $H_\parallel$) mode $\omega_L$, which has net moment oscillations *only* along the applied field direction (the *y*-axis). The high-energy mode corresponds to the transverse (with respect to $H_\parallel$) mode $\omega_T$, which has net moment oscillations in the *x-z* plane. The longitudinal mode $\omega_L$ drops to zero at the saturation field $H_S = 2H_E + H_A$.

The simple solution fits the experimental data well for the entire magnetic field range (dashed lines, Fig. 3d) with $H_A \approx 1.77$ T and $H_E \approx 0.76$ T. The interlayer exchange $H_E$ is in good agreement with the value from the spin-flip transition measurement under an out-of-plane field (Fig. 1c). The intralayer anisotropy $H_A$ or the saturation field ($H_S \approx 3.3$ T) is slightly smaller than the reported value [8], likely due to the different doping levels present in different samples (see gate dependence studies below). In contrast to the simple model, the measured $\omega_L$ is always finite likely due to the layer asymmetry in bilayer $CrI_3$ (caused by coupling to monolayer $WSe_2$), as well as inhomogeneous broadening (see below). The layer asymmetry also allows the observation of the low-frequency mode in the polar MOKE geometry, which would otherwise have zero out-of-plane magnetization.

Next we discuss the damping of the spin waves in 2D $CrI_3$. Figure 3e is the magnetic-field dependence of the normalized damping rate $\frac{2\pi}{\omega\tau}$ for both the transverse and longitudinal modes. Overall, damping is substantially higher below and near the saturation field for both modes. In addition, damping of the longitudinal mode is generally higher than the transverse mode. The high damping observed below and near $H_S$ is likely originated from inhomogeneous broadening of the magnetic resonances and spin wave dephasing. In this regime, the resonance frequencies are strongly dependent on internal magnetic interactions, which are sensitive to local doping and strain within the 2D layers. For instance, a ±10 % variation in the interlayer exchange field alone (which is comparable to the typical inhomogeneity reported in bilayer $CrI_3$ [3]) can account for the observed damping of the transverse mode at $H_S$. Inhomogeneous broadening also explains the seemingly larger damping for the longitudinal mode near $H_S$, where $\omega_L$ has a steep dependence on $H_E$ and $H_A$. Above $H_S$, the resonance frequencies are basically determined by the applied field and inhomogeneous broadening becomes insignificant, especially in the high-field limit (e.g. at 6 T). Other damping mechanisms such as interfacial damping and spin-orbit coupling of the iodine atom could become relevant here. However, our experiment on few-layer $CrI_3$ in the high-field limit shows weak dependence of $(\tau_T)^{-1}$ on layer number (Supplementary Fig. S6), suggesting that interfacial damping is not important. Future systematic studies are warranted to fully understand the microscopic damping mechanisms.

Finally we demonstrate control of the spin waves by electrostatic gating using a dual-gate device (Methods). Figure 4a shows the FFT amplitude spectra of coherent spin oscillations under a fixed magnetic field of 2 T at different gate voltages. The resonance shifts continuously from ~ 80 GHz to ~ 55 GHz when the gate voltage is varied from -13 V to +13 V (corresponding to from 'hole doping' to 'electron doping'). Figure 4b shows the entire magnetic-field dispersion of the transverse mode at varying gate voltages (the longitudinal mode is not studied because of its small amplitude). As in the zero gating case, the initial redshift of the mode is followed by a blueshift with increasing magnetic field at all gate voltages. The turning point, which is determined by the saturation field $H_S$, is tuned by about 1 T by gate voltage. Furthermore, while the dispersion of $\omega_T$ is nearly unchanged by gating above $H_S$, it is strongly modified below $H_S$.

In this regime the resonance frequency decreases by as much as 40% when the gate voltage is varied from -13 V to +13 V.

The observed magnetic-field dispersion of $\omega_T$ at all gate voltages can be described by the simple solution of the LLG equations discussed above (inset of Fig. 4b) with doping dependent interlayer exchange $H_E$ and intralayer anisotropy $H_A$ (Fig. 4c). Both fields decrease linearly with increasing gate voltage, with $H_A$ at a faster rate than $H_E$. A similar doping dependence for $H_E$ has been reported previously from the spin-flip transition measurement under an out-of-plane field [6]. Such doping dependences of the magnetic interactions can be understood as a consequence of doping dependent electron occupancy of the magnetic $Cr^{3+}$ ions and their wavefunction overlap. Based on this picture, increasing electron density weakens the magnetic interactions, and in turn the effective magnetic fields responsible for spin precession below $H_S$. Above $H_S$, the magnetization is fully saturated in the in-plane direction and the spin resonance frequency is almost solely determined by the applied field $H_\parallel$ and is therefore doping independent. A quantitative description of the experimental result, however, would require *ab initio* calculations and is beyond the scope of the current study.

In conclusion, we have demonstrated the generation and detection of spin waves in a prototype 2D magnetic material of bilayer $CrI_3$ with a time-resolved optical pump-probe method. The results allow the characterization of important parameters such as the internal magnetic interactions and damping. We have also demonstrated widely gate tunable magnetic resonances in this 2D magnetic system，revealing the potential of using 2D AFMs to achieve local gate control of spin dynamics for reconfigurable ultrafast spin-based devices [30,31].

**Methods**
**Sample and device fabrication**
The measured sample is a stack of 2D materials composed of (from top to bottom) few-layer graphite, hBN, monolayer $WSe_2$, bilayer $CrI_3$, hBN, and few-layer graphite. The top and bottom graphite/hBN pairs serve as gates. An additional stripe of graphite is attached to the $WSe_2$ flake for grounding and charge injection. The thickness of hBN layers is ~ 30 nm, and the graphite layers, about 2-6 nm. Bulk crystals of hBN were purchased from HQ graphene. Bulk $CrI_3$ crystals were synthesized by chemical vapor transport following methods described in previous reports[32,33]. These crystals crystallized into the *C*2/*m* space group with typical lattice constants of $a$ = 6.904 Å, $b$ = 11.899 Å, $c$ = 7.008 Å and $\beta$ = 108.74°, and Curie temperatures of 61 K. All layer materials were first exfoliated from their bulk crystals onto $SiO_2$/Si substrates and identified by their color contrast under an optical microscope. The heterostructure was built by the layer-by-layer dry transfer technique [34]. It was then released onto a substrate with pre-patterned gold electrodes, which contact the bottom gate, top gate, and grounding graphite flake. The steps involving $CrI_3$ before its full encapsulation in hBN layers were performed inside a nitrogen-filled glovebox because $CrI_3$ is air sensitive. In the gating experiment, equal top and bottom gate voltages were applied to the heterostructure and the gate voltage shown in Fig. 4 was the voltage on each gate.

**Time-resolved magneto-optical Kerr effect (MOKE) and magnetic circular dichroism (MCD)**

In the time-resolved MOKE setup, the probe beam is the output of a Ti:Sapphire oscillator (Coherent Chameleon with a repetition rate of 78 MHz and pulse duration of 200 fs) centered at 1.54 eV, and the pump beam is the second harmonic of an optical parametric oscillator (OPO) (Coherent Chameleon compact OPO) output centered at 1.73 eV. The time delay between the pump and probe pulses was controlled by a motorized linear delay stage. Both the pump and probe beam were linearly polarized. The pump intensity was modulated at 100 kHz by a combination of a half-wave photoelastic modulator (PEM) and a linear polarizer whose transmission axis is perpendicular to the original pump polarization. The pump and probe beam impinged on the sample at normal incidence. The reflected light was first filtered to remove the pump, passed through a half-wave Fresnel rhomb and a Wollaston prism, and detected by a pair of balanced photodiodes. The pump-induced change in Kerr rotation was determined as the ratio of the intensity imbalance of the photodiodes obtained from a lock-in amplifier locked at the pump modulation frequency and the intensity of each photodiode.

For the MCD measurements, a single beam centered at 1.8 eV was used. The light beam was modulated at 50 kHz between the left and right circular polarization using a PEM. The reflected light was focused onto a photodiode. The MCD was determined as the ratio of the ac component of the photodiode signal measured by a lock-in amplifier at the polarization modulation frequency and the dc component of the photodiode signal measured by a voltmeter.

For all measurements samples were mounted in an optical cryostat (attoDry2100) with a base temperature of 1.7 K and a superconducting solenoid magnet up to 9 Tesla. For measurements under an out-of-plane field, the sample was mounted horizontally and light was focused onto the sample at normal incidence by a microscope objective. For measurements under an in-plane field, the sample was mounted vertically and the light beam was guided by a mirror at 45° and focused onto the sample at normal incident with a lens.

**Landau-Lifshitz-Gilbert (LLG) equations**
We model the field dependent spin dynamics in antiferromagnetic bilayer CrI$_3$ using coupled Landau-Lifshitz-Gilbert (LLG) equations [29],

$$\frac{\partial \boldsymbol{M}_i}{\partial t} = -\gamma \boldsymbol{M}_i \times \boldsymbol{H}_i^{eff} + \frac{\alpha}{M_S} \boldsymbol{M}_i \times \frac{\partial \boldsymbol{M}_i}{\partial t}. \qquad (1)$$

where $i = 1, 2$. In Eqn. 1 $\boldsymbol{M}_i$ is the magnetization of the top or bottom layer (which are assumed to have an equal magnitude $M_S$), $\gamma/2\pi \approx 28$ GHz/T is the electron gyromagnetic ratio, $\alpha$ is the dimensionless damping factor, and $\boldsymbol{H}_i^{eff}$ is the effective magnetic field in each layer that is responsible for spin precession. In the absence of applied magnetic field, $\boldsymbol{M}_1$ and $\boldsymbol{M}_2$ are anti-aligned along the easy axis (z-axis). When an in-plane field $\boldsymbol{H}_\parallel$ (along the y-axis) is applied, $\boldsymbol{M}_1$ and $\boldsymbol{M}_2$ are tilted symmetrically towards the y-axis, before fully turned into the applied field direction at the saturation field $H_S = 2H_E + H_A$. Here $H_E$ and $H_A$ are the interlayer exchange and intralayer anisotropy fields, respectively. A schematic is shown in Fig. 3c. The effective field $\boldsymbol{H}_{1,2}^{eff} = \boldsymbol{H}_\parallel - \frac{H_E}{M_S}\boldsymbol{M}_{2,1} + \frac{H_A}{M_S}(\boldsymbol{M}_{1,2})_z \hat{\boldsymbol{z}}$ has contributions from the applied field, the interlayer exchange field, and the intralayer anisotropy field. We search for solution in the form of a harmonic wave $e^{i\omega t}$ with angular frequency $\omega$. For the simple case of zero damping ($\alpha = 0$), two eigenmode frequencies $\omega_T$ and $\omega_L$ are given in the main text.

In case of finite but weak damping, we find the following transverse and longitudinal modes after simplifying the LLG equations:

Before saturation ($H_\parallel < H_S$),

$$\omega_T^2(1+\alpha^2) - i\alpha\omega_T\gamma\left(\frac{\omega_{T0}^2/\gamma^2}{2H_E+H_A} + 2H_E + H_A\right) - \omega_{T0}^2 = 0;$$

$$\omega_L^2(1+\alpha^2) - i\alpha\omega_L\gamma\left(\frac{\omega_{L0}^2/\gamma^2}{H_A} + H_A\right) - \omega_{L0}^2 = 0;$$

After saturation ($H_\parallel > H_S$),

$$\omega_T^2(1+\alpha^2) - i\alpha\omega_T\gamma(2H_\parallel - H_A) - \omega_{T0}^2 = 0;$$

$$\omega_L^2(1+\alpha^2) - i\alpha\omega_L\gamma(2H_\parallel - 4H_E - H_A) - \omega_{L0}^2 = 0.$$

Here $\boldsymbol{\omega_{T0}}$ and $\boldsymbol{\omega_{L0}}$ correspond to the solution at zero damping ($\boldsymbol{\alpha} = 0$). In particular, when $\boldsymbol{\alpha} \ll 1$, the oscillation frequency (the real part of the solution for $\boldsymbol{\omega_T}$ and $\boldsymbol{\omega_L}$) becomes $\frac{\omega_0}{\sqrt{1+\alpha^2}}$, where $\boldsymbol{\omega_0}$ is the undamped solution for the two modes. Overall, the eigenmode frequencies are reduced due to damping, and the two modes will no longer be degenerate at $\boldsymbol{H_\parallel = 0}$ taking into account of higher order corrections of $\boldsymbol{\alpha}$. At low temperature, we found this correction insignificant for the high-frequency branch, which has a larger oscillation amplitude and was measured with a higher precision. Fitting the experimental data with the damped LLG solution yielded similar values for $\boldsymbol{H_E}$ and $\boldsymbol{H_A}$.

**Mechanism for ultrafast excitation of coherent magnons**
We have investigated the mechanism for the observed ultrafast excitation of magnons in bilayer CrI$_3$. A plausible picture involves exciton generation in WSe$_2$ by the optical pump, ultrafast exciton dissociation and charge transfer at the CrI$_3$-WSe$_2$ interface, and an impulsive perturbation to the magnetic anisotropy and exchange fields in CrI$_3$ by the injected hot carriers. Several control experiments were performed to test this picture. Pump-probe measurements were performed on both monolayer WSe$_2$ and bilayer CrI$_3$ areas alone (non-overlapped regions in the heterostructure) under the same experimental conditions. Negligible pump-induced MOKE signal was observed. In addition, measurement was done on the heterostructure at different pump energies. The magnetic resonance frequencies were found unchanged, but the amplitudes follow the absorption spectrum of WSe$_2$ (Supplementary Fig. S1). These two experiments show that magnons are generated through optical excitation of excitons in WSe$_2$. It has been reported earlier that CrI$_3$-WSe$_2$ heterostructures have a type-II band alignment, which can facilitate ultrafast exciton dissociation and charge transfer [26]. Next the onset of coherent oscillations is instantaneous with optical excitation in our experiment. This excludes lattice heating in CrI$_3$ as a dominant mechanism for the generation of magnons, which typically takes a longer time to build up. Moreover, the resonance amplitude is independent of the pump laser polarization (Supplementary Fig. S2), indicating that hot carriers, rather than the angular momentum of the carriers, are responsible for the excitation of magnons. Finally, as we show in the main text, the magnetic anisotropy and exchange can be effectively altered by carrier doping in CrI$_3$. These experiments are all consistent with the proposed mechanism of ultrafast excitations of magnons in CrI$_3$-WSe$_2$ heterostructures.

**Temperature dependence of magnon modes**

We have performed the optical pump/MOKE probe experiment in CrI$_3$-WSe$_2$ heterostructures at temperature ranging from 1.7 K to 50 K. No obvious oscillations can be measured above 50 K when bilayer CrI$_3$ is close to its Néel temperature. The results at 1.7 K are presented in the main text. Supplementary Fig. S3 and S4 show the corresponding measurements and analysis for 25 K and 45 K, respectively. With increasing temperature, the magnon frequency decreases and the saturation field (estimated from the minimum of the frequency dispersion) also decreases. A systematic temperature dependence is shown in Supplementary Fig. S5 for the high-frequency mode $\omega_T$ at a fixed in-plane field of 2 T. The frequency has a negligible temperature dependence well below the Néel temperature (< 20 K), and decreases rapidly when the temperature approaches the Néel temperature.

**Additional measurements on few-layer CrI$_3$**

We have measured the magnetic response from a few-layer CrI$_3$ (6-8 layer) sample. Because of the larger MOKE signal and higher optical absorption in thicker samples, magnetic oscillations can be measured without the enhancement from monolayer WSe$_2$. The results are shown in Supplementary Fig. S6. The comparison of results from samples of different thicknesses provides insight into the origin of magnetic damping. For instance, in the high-field limit (6 T), few-layer and bilayer CrI$_3$ show a similar level of damping. This indicates that interfacial damping is not the dominant contributor to damping.

**Competing interests**

The authors declare no competing interests.

**Data availability**

The data that support the findings of this study are available within the paper and its Supplementary Information. Additional data are available from the corresponding authors upon request.

**Figures and figure captions**

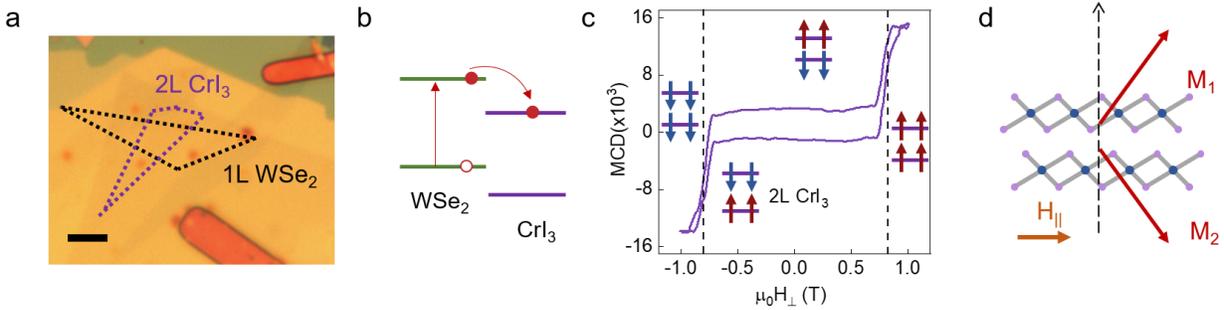

**Figure 1 | Bilayer CrI$_3$/monolayer WSe$_2$ heterostructure. a,** Optical microscope image of the heterostruture. Bilayer CrI$_3$ is outlined with a purple line, and monolayer WSe$_2$, a black line. Scale bar is 5 $\mu$m. **b,** Schematic of a type-II band alignment between monolayer WSe$_2$ and CrI$_3$. Optically excited exciton in WSe$_2$ is dissocated at the interface and electron is transferred to CrI$_3$ [26]. **c,** MCD of the heterostrucutre as a function of out-of-plane magnetic field at 4 K. Hysteresis is observed for field sweeping along two opposing directions. Insets are schematics of the corresponding magnetizations in the top and bottom layers of blayer CrI$_3$. The dashed lines indicate the spin-flip transition around 0.75 T. **d,** Schematic of bilayer CrI$_3$ under an in-plane magnetic field $H_\parallel$. Below the saturation field, the magnetizations of the top and bottom layer are symmetrically canted towards the applied field direction.

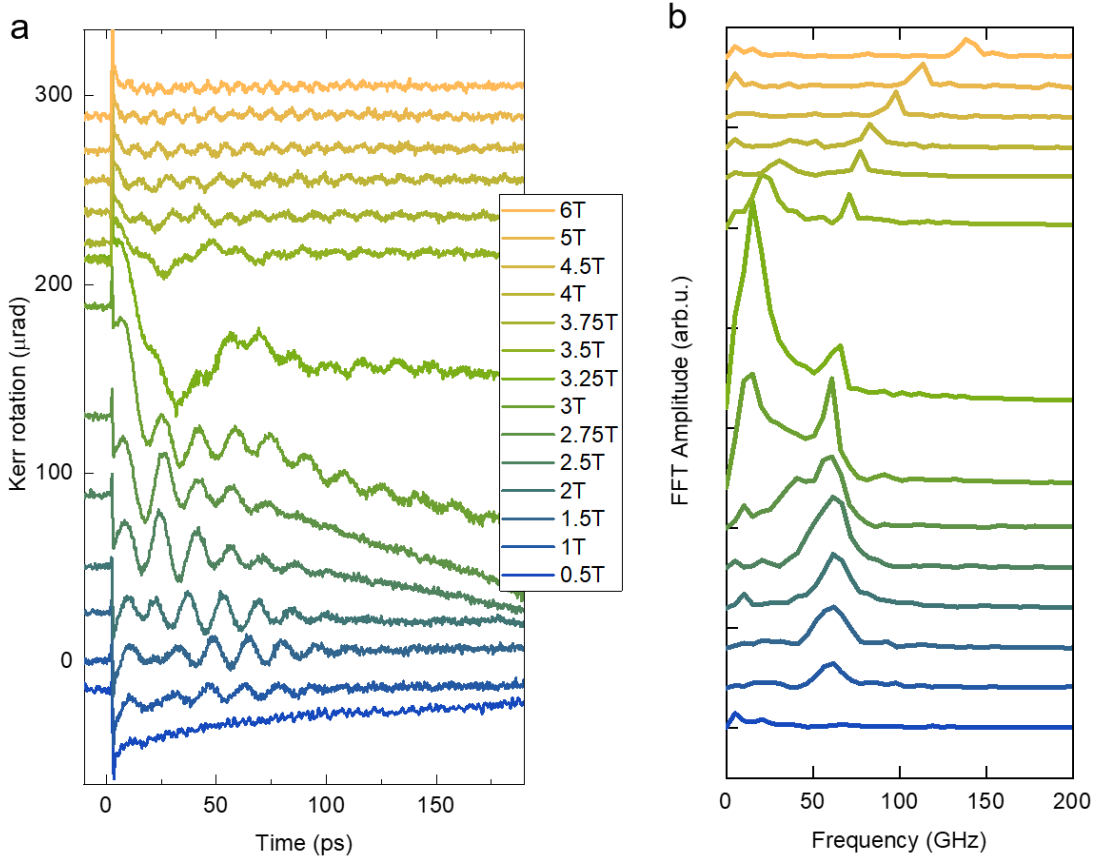

**Figure 2 | Time-resolved magnon oscillations. a,** Pump-induced Kerr rotation as a function of pump-probe delay time in bilayer $CrI_3$ under different in-plane magnetic fields. The curves are displaced vertically for clarity. **b,** FFT amplitude spectra of the time dependences shown in **a** after the demagnetization dynamics (exponential decay) were removed. The spectra are vertically displaced for clarity.

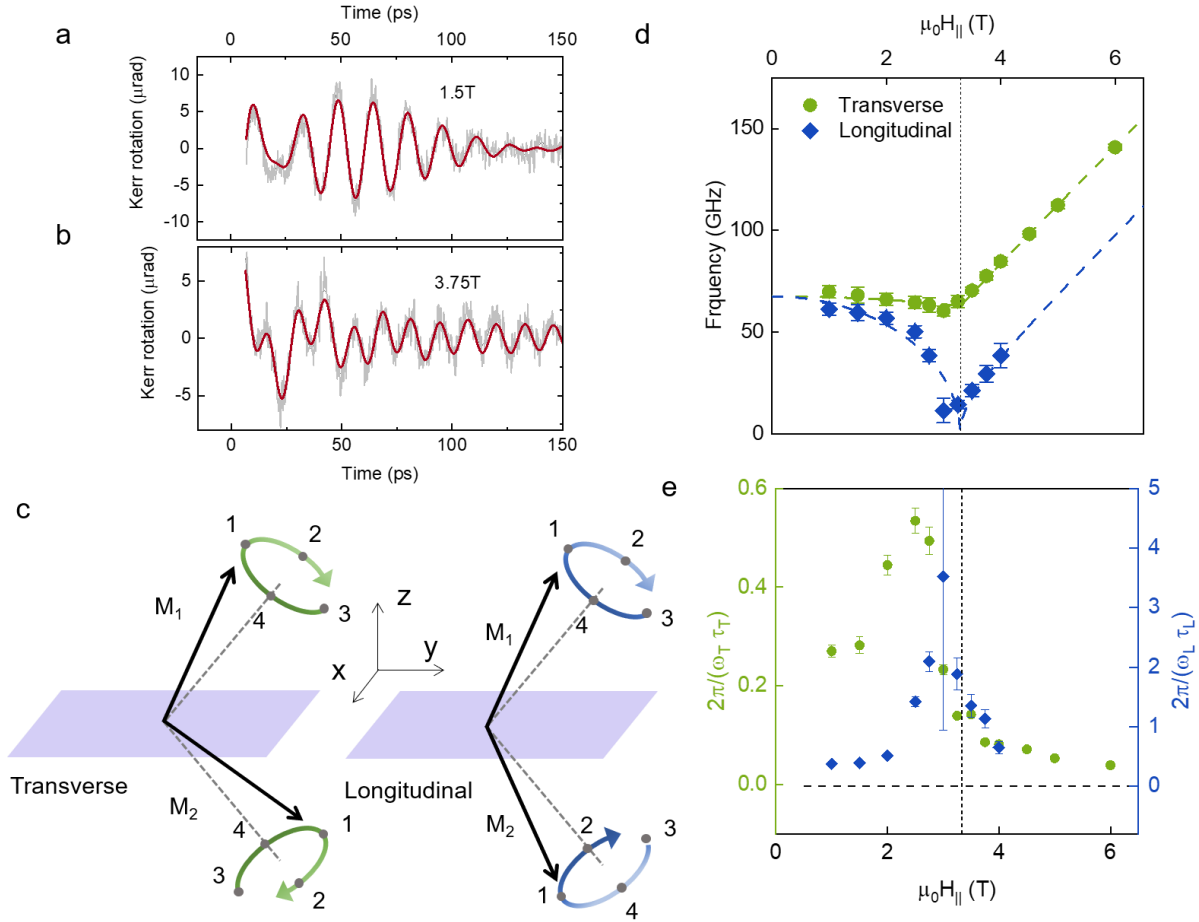

**Figure 3 | Magnon dispersion and damping. a, b,** Pump-induced MOKE dynamics in bilayer CrI$_3$ under two representative in-plane fields of 1.5 T (**a**) and 3.75 T (**b**). Grey lines are experiment after subtracting the demagnetization dynamics, and red lines, fits to two damped harmonic oscillations. **c,** Illustration of two spin wave eigenmodes in an AFM: the transverse mode (left) and the longitudinal mode (right). The dashed lines indicate the equilibrium top and bottom layer magnetization M$_1$ and M$_2$, which are titled symmetrically from the *z*-direction towards the applied field direction (*y*-axis). The magnetizations precess following the green and blue arrows in the order 1 through 4. **d, e,** Oscillation frequencies (**d**) and damping rates (**e**) of the transverse and longitudinal modes extracted from the two harmonic oscillation fit as a function of in-plane magnetic field. The error bars are the fit uncertainties. The vertical dotted lines indicate the in-plane saturation magnetic field. Dashed lines in **d** are fits to the LLG equations as described in the text.

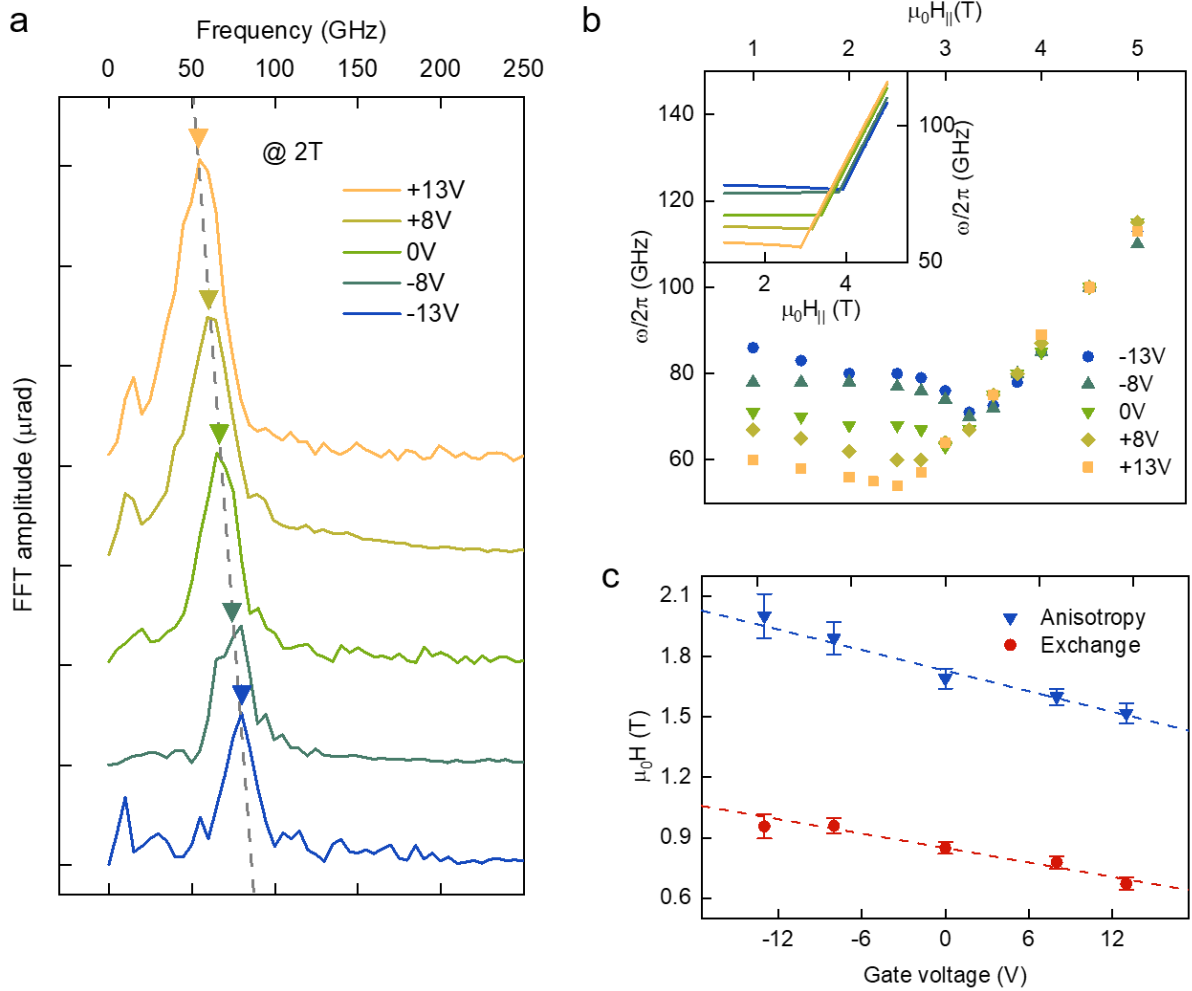

**Figure 4 | Gate tunable magnon frequency. a,** FFT amplitude spectra of the magnons as a function of gate voltage under a fixed in-plane field of 2 T. The dashed line is a guide to the eye of the evolution of the resonance frequency with gate voltage and triangles indicate the peak of the resonance. **b,** Magnetic-field dispersion of the transverse mode at different gate voltages. The inset shows the fits of the experimental data to the LLG equations. The same colored lines (LLG equation) and symbols (experiment) denote the same gate voltage. **c,** Anisotropy field $H_A$ and exchange field $H_E$ extracted from the fits in **b** at different gate voltages. Error bars are the standard deviation from the fitting. Dashed lines are linear fits.

**Supplementary figures**

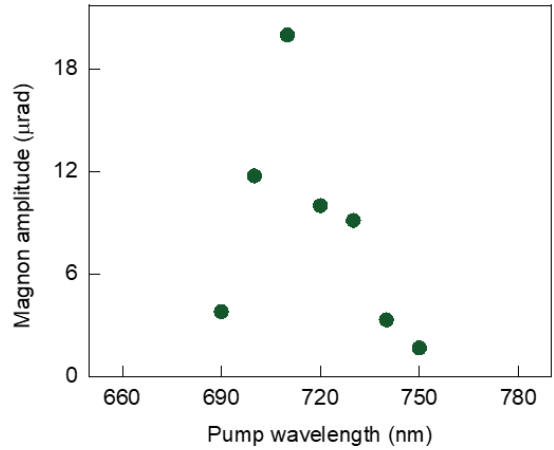

**Figure S1 |** Amplitude of the spin waves under a fixed in-plane magnetic field of 2 T as a function of pump wavelength. The dependence resembles that of the excitonic resonance in monolayer $WSe_2$. The spectral broadening arises from the additional $WSe_2$ trion absorption and the linewidth of the light pulses (~ 5 nm in full width at half maximum (FWHM)) employed in the pump-probe measurement.

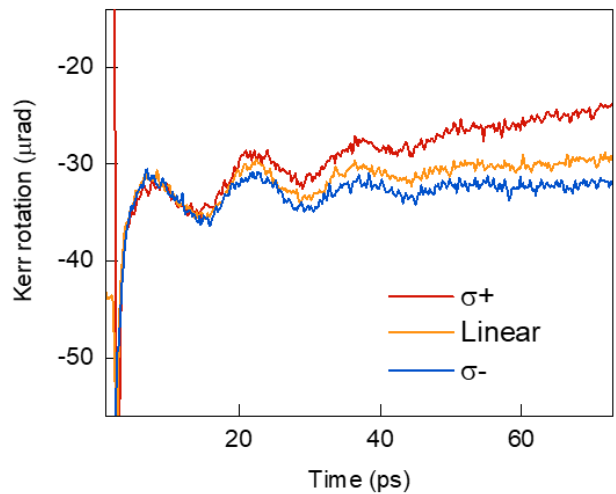

**Figure S2 |** Spin wave dynamics under $H_{\parallel}$ = 2 T excited by optical pump of different polarizations. The red, orange and blue lines correspond to left circularly polarized, linear polarized, and right circularly polarized pump, respectively. The curves were vertically shifted for easy comparison. The oscillation amplitude does not depend on the pump polarization (i.e. photon angular momentum).

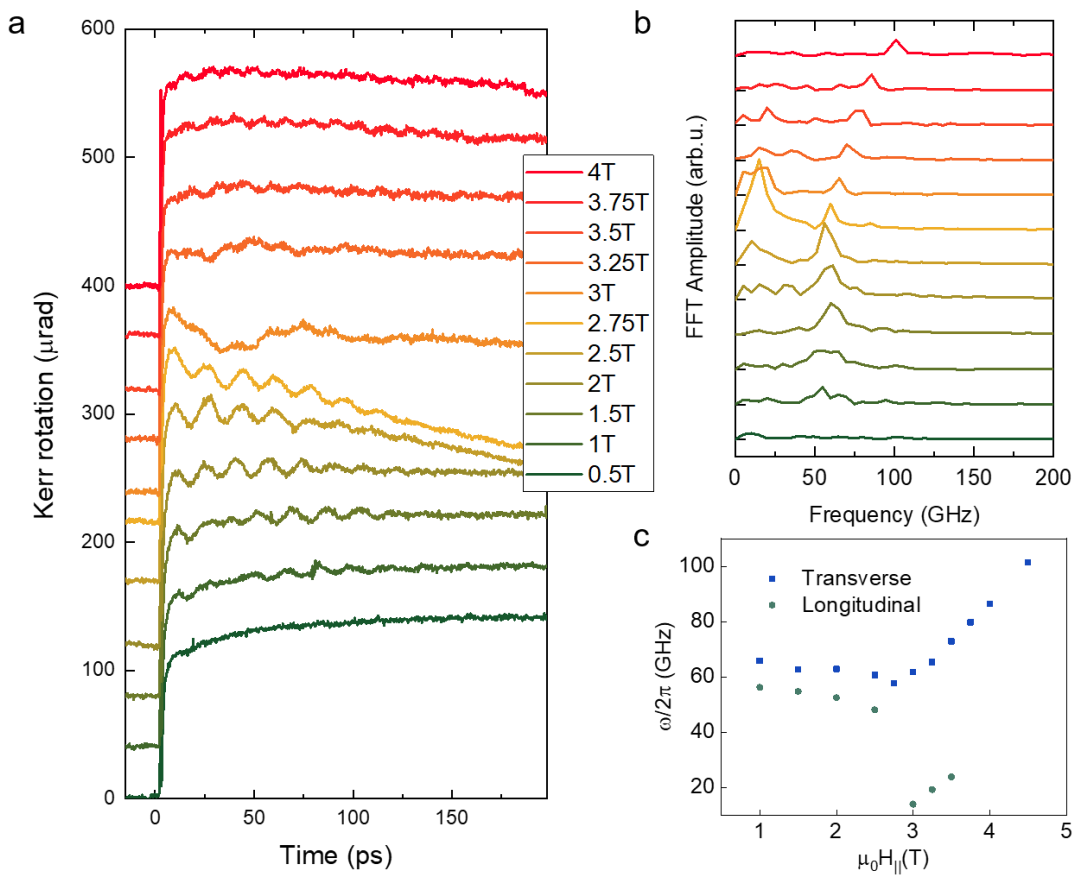

**Figure S3 | Magnon oscillations at 25 K. a,** Spin dynamics in bilayer CrI$_3$ under different magnetic fields. The curves were vertically displaced for clarity. **b,** FFT amplitude spectra of **a. c,** In-plane field dispersion of the two magnon modes extracted from fitting the time-resolved MOKE signal with two harmonic oscillations.

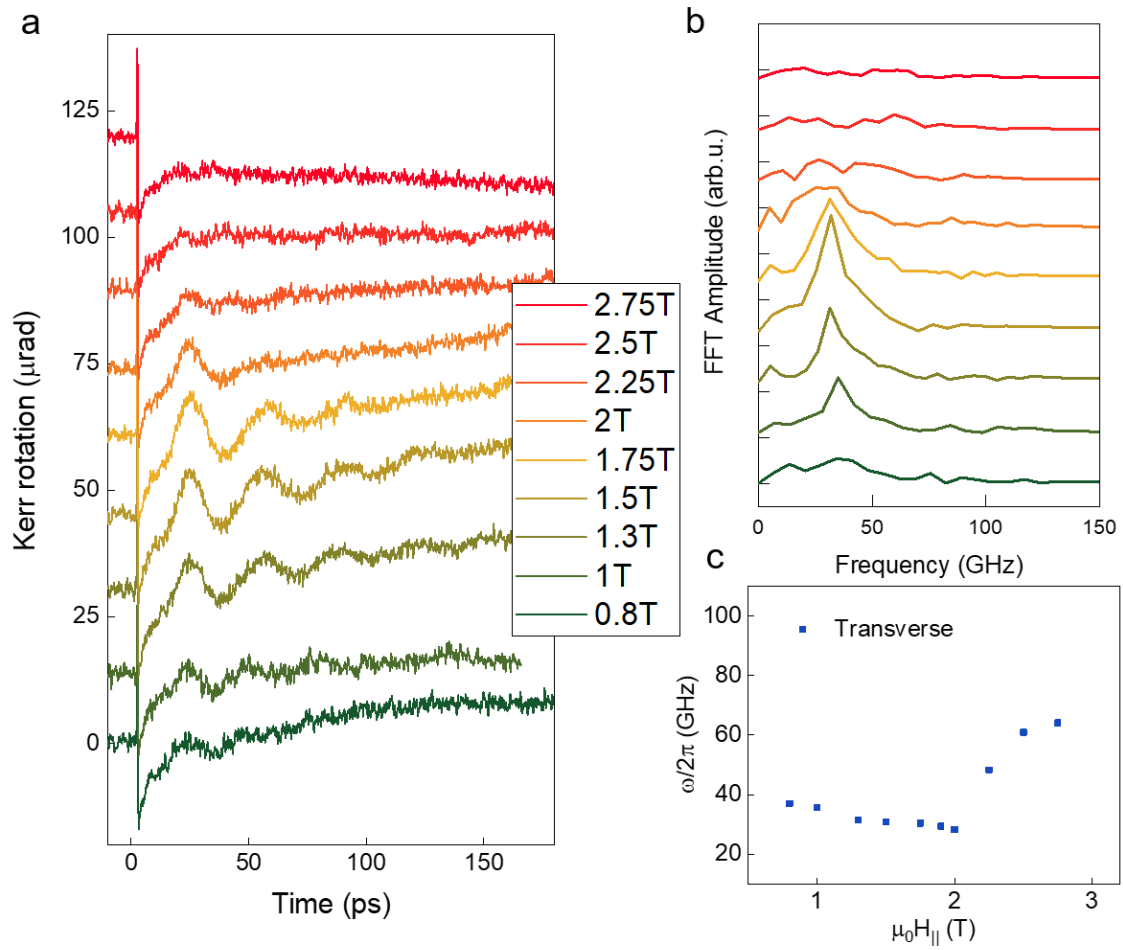

**Figure S4 | Magnon oscillations at 45 K.** Same as in Supplementary Fig. S3. Due to the weak signal, we can only identify the transverse mode $\omega_T$.

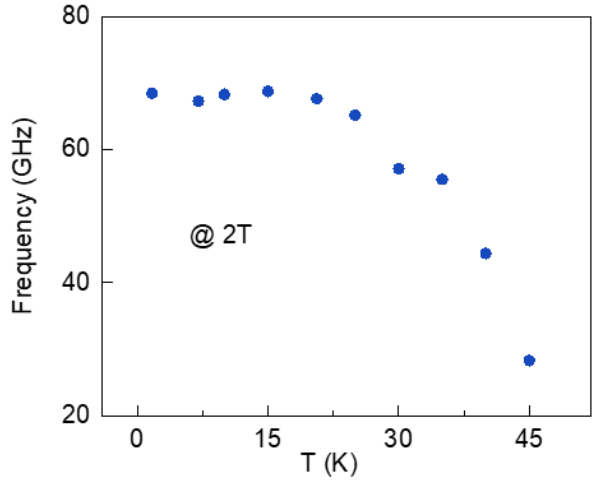

**Figure S5 |** Temperature dependence of the transverse magnon mode frequency under a fixed in-plane magnetic field of 2 T. All other experimental conditions are the same as in Fig. S3 and S4.

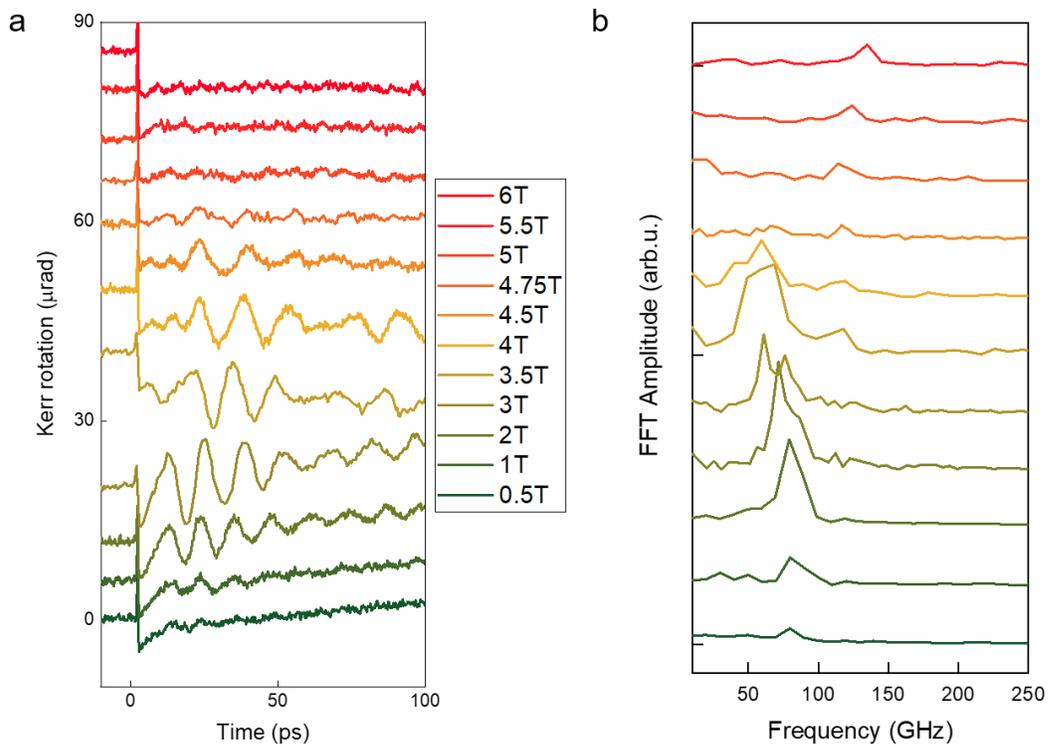

**Figure S6 | Pump-probe measurements on few-layer CrI$_3$ at 1.7 K. a,** Spin wave dynamics under different in-plane magnetic fields. **b,** The corresponding FFT amplitude spectrum of **a**. The damping at 6T is estimated to be ~0.04, which is similar to bilayer CrI$_3$.

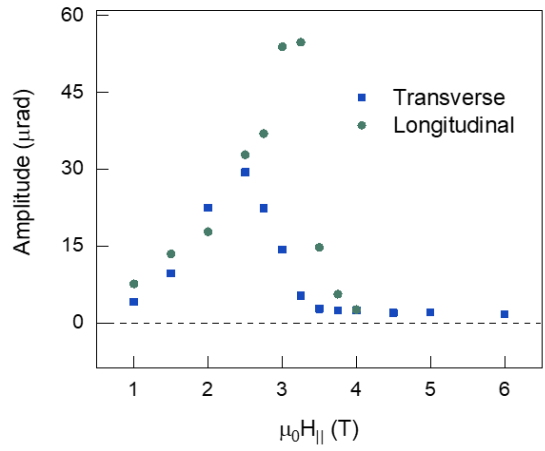

**Figure S7 |** Amplitude of the longitudinal and transverse magnon modes extracted from the time-resolved MOKE measurement (Fig. 2 of the main text).